\documentstyle[prd,aps,preprint]{revtex}
\tightenlines
\newcommand{\be}{\begin{equation}}
\newcommand{\ee}{\end{equation}}
\newcommand{\bea}{\begin{eqnarray}}
\newcommand{\beas}{\begin{eqnarray*}}
\newcommand{\eea}{\end{eqnarray}}
\newcommand{\eeas}{\end{eqnarray*}}
\newcommand{\ba}{\begin{array}}
\newcommand{\ea}{\end{array}}

\begin{document}

\draft
\preprint{\vbox{
\hbox{UMD-PP-00-065}}}

\title{Cosmology of Brane-Bulk Models in Five Dimensions}

\author{ R. N. Mohapatra$^1$\footnote{e-mail:rmohapat@physics.umd.edu},
A. P\'erez-Lorenzana$^{1,2}$\footnote{e-mail:aplorenz@glue.umd.edu} 
and C. A. de S. Pires$^{1}$\footnote{e-mail:cpires@physics.umd.edu}}

\address{
$^1$ Department of
Physics, University of Maryland, College Park, MD, 20742, USA\\
$^2$  Departamento de F\'\i sica,
Centro de Investigaci\'on y de Estudios Avanzados del I.P.N.\\
Apdo. Post. 14-740, 07000, M\'exico, D.F., M\'exico. }

\date{March, 2000}

\maketitle 

\begin{abstract} 
We study the  cosmology of models with four space and one time dimension 
where our universe is a 3-brane and report a few results which extend
existing work in several directions.  Assuming a stable fifth 
dimension, we obtain a solution for the metric, which does not depend on 
any arbitrary parameters. We discuss some implications of this result.
\end{abstract}

\vskip0.5in

\section{Introduction}

Over the last two years there has been a lot of interest in
models  where our universe is a 3-brane (a hyper-surface) embedded in a
higher dimensional bulk. The standard model particles are confined to the
brane whereas gravity propagates in the bulk. Such models were conjectured
early on~\cite{rubakov}
as interesting possibilities and have recently been argued as plausible
solutions of type I string theories~\cite{witen}. One of the attractive
features of these models is the intriguing possibility that the
fundamental scale, $M$, identified
as the string scale,  could be lower than the Planck scale, 
$M_{P\ell}=(8\pi G_N)^{-1/2}$, by several orders of magnitude~\cite{early},
perhaps even of TeV range~\cite{dvali}. This last possibility 
may provide a new way to solve the hierarchy problem between the
electroweak scale and the scale of
gravity i.e. both scales being of same order, there is no hierarchy to
worry about. The large value of the Planck scale in this picture owes its
origin to the existence of very large hidden extra dimensions in
nature. The price one has to pay is of course that now one
has to understand why the extra dimension(s) is(are) so large. 
The relation between  the fundamental scale, the $M_{P\ell}$ 
and the volume of the external space, $V_n$, is
given by~\cite{dvali}
 \be 
 M^{2+n} V_n = M^2_{P\ell}. 
 \ee 
This picture leads to a modification of the  inverse square law of
gravity at small distances, $r\sim V_n^{1/n}$ and can therefore be probed
experimentally. If $M$ is in the TeV range, string theories become
accessible to collider tests. All these make the idea
phenomenologically quite attractive. It is therefore interesting to study
the cosmology of these models.

One of the first things that one
needs to know in order to study the cosmology of these models is the time
dependence of the metric. In addition, the
metric will also have a dependence on the bulk coordinate $y$, even when
the bulk is totally empty, simply because the presence of a brane induces 
a nonzero curvature. It turns out that 
 this fact leads to a variety of interesting consequences for the
cosmology of  such 
models~\cite{gog,rs,rs2,binetruy1,waldram,cline,olive,binetruy2,shiru,more}.
In particular, the new bulk-brane  picture seems to drastically
modify the standard time evolution law in the brane-confined universe. The
five dimensional Einstein equations first studied in \cite{binetruy1} 
 implies that the Hubble parameter
$H$ is proportional to the density on the brane, $\rho$,
instead of the usual $H\sim\sqrt\rho$ of the standard big bang cosmology.
Since the successes of the standard cosmology such as
nucleosynthesis and the common understanding of subsequent evolution rely
crucially on the assumption of $H\sim\sqrt\rho$,
a great deal of work has been devoted to understand how this standard
behaviour of $H$ can be recovered. Some
ideas proposed to solve this problem include
cancellation of bulk and brane cosmological
constants~\cite{cline}, consideration of a non zero thickness of
the brane~\cite{olive} etc.  It has also been noted that  Friedmann
equation could be recovered on the basis of  fine tuning of cosmological
constants even when
the extra dimension is not stable~\cite{binetruy2} and no matter
in the bulk. However, in
the process  it has been found that a free parameter $\cal C$, a
constant of integration,  appears in this equation. It contributes to the
evolution of the
Hubble parameter in the form of an effective radiation term, jeopardizing
the cosmic scale parameters.  The same arbitrary parameter has been
used to study cosmological evolution even when the bulk radius is stable
~\cite{binetruy2}. We reexamine this in this paper.
Writing the equations governing the evolution in time, $t$ and the
fifth coordinate, $y$ of the various parameters defining the metric (the
analogs of the Friedman-Robertson-Walker (FRW) scale factor), we show
that it is possible to get a solution for the metric as a function of the
bulk coordinate without any arbitrary constant, if the bulk radius 
is stable. This is the main result of this brief note. We then point out
some of the implications of this result.

This paper is arranged as follows: in section 2, we review 
Einstein equations in a diagonal metric, commonly used in
the literature; in section 3, we look for
solutions of Einstein
equation keeping the bulk radius stable and obtain an explicit form
for the metric which involves only the known FRW scale factor and no
extra parameters. We recover the known behaviour for the Hubble parameter,
i.e. $H$ depends linearly on the brane energy density. We also
show that the
equation involving the brane-space like components can be derived from
the other equations;
they reduce to the acceleration equation for the FRW scale factor.
Finally, we reconsider some examples already studied in \cite{binetruy2}
and rederive the  exact solutions for the metric in the prescence of a
cosmological constant in the bulk. In section 5, we make some remarks
on the general nonfactorizable  nature of this solutions.

\section{Basic framework}
\setcounter{equation}{0}

The basic framework for our discussion is a five dimensional
space-time where a flat (zero thickness) brane is localized at the
position identified as $y=0$ along the fifth dimension.
Although in this
paper, we assume that the extra dimension is compact, this is
not crucial for our conclusions and  our results also apply when
the extra dimension is noncompact.

Since we are interested in the brane cosmology, we start by adopting
the cosmological principle of isotropy and homogeneity in the three space
dimensions of the brane. The presence of the brane clearly breaks the
isotropy along the fifth dimension and this is reflected 
in the explicit $y$ dependence of the metric tensor which we choose to
have the following form:
 \be
  ds^2=-n^2 (y,t) dt^2 + a^2(y,t)\gamma_{ij}dx^i dx^j + b^2(y,t) dy^2.
  \label{4}
 \ee
Here $\gamma_{ij} = f(r) \delta_{ij}$, with $f^{-1}(r)=1-kr^2$ being the
usual Robertson-Walker curvature term, where $k=-1,0,1$; $t$ and $x^i$,
$i=1,2,3$ are the time and space-like coordinates along the brane
respectively.

The five dimensional Einstein equations take the form
 \be
 G_{AB} = R_{AB}-\frac{1}{2} g_{AB} R = 
 \kappa_5^2 \left[\hat T_{AB} + 
 T_{\mu\nu}\ \delta^\mu_A\ \delta^\nu_B\ \delta(b y)\ \right],
 \label{1}
 \ee
where $\kappa_5^2=8\pi G_{(5)}=M^{-3}$ is the five dimensional coupling
constant of gravity, $R_{AB}$ is the five dimensional Ricci tensor and $R$
the scalar curvature, $A,B= 0,1,2,3,4$ and $\mu,~\nu= 0,1,2,3$. In
conformity with the usual practice, we identify
the mass parameter $M$ with
the string scale. In the last expression the various source terms have
been explicitly separated.  $T_{\mu\nu}$ and ${\hat T}_{AB}$ represent the 
stress-energy-momentum tensors of  the brane and bulk respectively.  For
the scenarios we are going to discuss from now on, it is sufficient to
work in the perfect fluid approximation to those tensors
 \bea
 {\hat T}^A{_B} &=& diag(-\rho_B,P_B,P_B,P_B,P_T),\nonumber \\
 {T^\mu}_\nu &=& diag(-\rho_b,p_b,p_b,p_b).
 \label{3}
 \eea 
where $\rho_B$, $P_B$ and $P_T$ represent the densities and pressure on
the bulk and, respectively, $\rho_b$ and  $p_b$ are those on the
brane. Notice that assuming   $\hat T_{04}=0$ avoids the
complication  of a matter flux along the fifth dimension. 

So far, most of the attention paid to this model in literature appears to
have focussed on the case where vacuum
energy (a cosmological constant) is the only component of the bulk stress
tensor. In general however, all the components of $T^A_B$ could be 
functions of time, if they are in the brane
and could depend on both $t$ and $y$ if they describe the
bulk~\cite{olive}. Therefore, in our analysis, we will keep the 
energy-momentum tensor dependent on $t$ (and $y$ for the bulk terms).
It is of course straightforward to see that more branes could be
considered by including their
corresponding stress tensors in  Eq. (\ref{1}) and our
discussion below easily generalises to this case.

In order to solve the Einstein equations on the presence of the
delta-function type
densities, we proceed as follows. First we observe that the 
brane divides the bulk
into two different domains, where the only source is $\hat T_{AB}$. We
then solve the equations in each domain separately, and impose the
boundary conditions at the brane to get the global
solution. This also helps to define the metric on the brane itself.
First, the metric tensor, $g$, clearly should be continuous, i.e. the
solutions must satisfy
 \be
  g_{AB}(y=0^-)=g_{AB}(y=0^+).
 \ee
Next, as has already been noted, since $G_{AB}$
involves  up to  second derivatives on the metric tensor  with respect of
$y$, we must use them to match the delta function
distributions~\cite{binetruy1}.  Technically speaking this means
that the extrinsic curvature $K_{AB}$ in  the Gauss-Codacci
formulation~\cite{shiru}
should be discontinuous at the position of the branes. Then, by
integrating Eq. (\ref{1}) at both sides of the brane, we will get matching
conditions for the first derivatives.  This leads to the constraint
 \be 
  \int^{0+}_{0^-} \! dy~ b~ G_{\mu\nu} = \kappa_5^2 T_{\mu\nu}.
  \label{int}
 \ee 
 To evaluate this integral we should assume that all other terms not
involving second derivatives on $y$ are finite. 

Typically, a parity symmetry $P: y\rightarrow -y$ is assumed as in the
Horava-Witten model~\cite{witen}. Physically, this symmetry could be seen
as  a residual effect of the broken isotropy along the fifth dimension,
as in the $S^1/Z_2$ orbifold construction of ref.\cite{witen}.
We will  assume this hereafter. In the presence of other
branes that explicitly break this  symmetry,
our discussions have to be reconsidered depending on the number of branes.

Let us now proceed to the details.  Using the above form of the bulk
stress tensor, the non trivial components of the Einstein equations (away
from the brane) are given as~\cite{binetruy1}
 \bea
  G_{00}&=& 
   3\left\{ \frac{\dot a}{a} \left( \frac{\dot a}{a}+
   \frac{\dot b}{b}\right)-
   \frac{n^2}{b^2} \left[ \frac{a'}{a}
   \left(\frac{a'}{a} - \frac{b'}{b} \right)  + \frac{a''}{a}\right] +
   k \frac{n^2 }{a^2 } \right\} = \kappa^2_5 n^2 \rho_B, 
   \label{G00}\\
  G_{ij}&=&
   \frac{a^2}{b^2} \left\{ \frac{a'}{a} \left( 2\frac{n'}{n} + 
   \frac{a'}{a} \right) - \frac{b'}{b}\left(\frac{n'}{n} + 
   2\frac{a'}{a}\right) + 2\frac{a''}{a} + 
   \frac{n''}{n}\right\}\gamma_{ij} + \nonumber \\
   && \frac{a^2}{n^2} \left\{\frac{\dot a}{a} 
   \left(2\frac{\dot n}{n}-\frac{\dot a}{a}\right)  + 
   \frac{\dot b}{b} \left( \frac{\dot n}{n}- 2\frac{\dot a}{a}\right) -
   2\frac{\ddot a}{a} - 
   \frac{\ddot b}{b}\right\}\gamma_{ij} -  
   k\gamma_{ij}=\kappa^2_5 a^2 P_B\gamma_{ij},
    \label{Gij}\\
  G_{04}&=&
   3\left(\frac{n'}{n}\frac{\dot a}{a}+\frac{a'}{a}\frac{\dot b}{b}
   -\frac{\dot a'}{a}\right)=0, 
   \label{G04}\\
  G_{44}&=&
   3\left\{\frac{a' }{a}\left(\frac{a' }{a}+\frac{n' }{n}\right)-
   \frac{b^2}{n^2}\left[\frac{\dot a}{a} \left(\frac{\dot a}{a}-
   \frac{\dot n}{n}\right) + \frac{\ddot a}{a}\right] - 
   k\frac{b^2}{a^2}\right\}=\kappa^2_5  b^2 P_T ;
   \label{G44} 
 \label{8}
 \eea 
where primes (dots) are used to denote derivatives with respect to $y$
($t$).
This system of equations is supplemented by the Bianchi identity
$\hat T^A_{~~B;A}=0$,
which translates into the conservation laws~\cite{olive}
 \bea
  \dot\rho_B + 3 {\dot a \over a} \left( \rho_B + P_B \right) + 
  {\dot b \over b}\left( \rho_B + P_T \right) &=& 0 
 \label{drho}\\
  P'_T + 3 {a' \over a} \left( P_T - P_B\right) + 
  {n' \over n} \left( P_T + \rho_B\right) = 0
 \label{dp}
 \eea

Next, by using Eq. (\ref{int}), the following boundary conditions are
easily obtained~\cite{binetruy1}
\bea
  {\Delta a'\over a b }\bigg|_{0} &=& -{\kappa_5^2\over 3}\rho_b, 
  \label{gap1}\\
  {\Delta n'\over nb }\bigg|_{0} &=& {\kappa_5^2\over 3}(3p_b + 2\rho_b);
   \label{gap2}
 \eea 
where the left hand side of the above equations has to be evaluated at the
position of the brane,  and the function  $\Delta a'(0):=
a'(0^+) - a'(0^-)$ give  the size of the jump of the derivative of
$a(y)$. The same applies for $\Delta n'$. Since we are assuming $P$ parity, 
the jump on the above equations could be expressed in terms of the
limiting value on one side of the brane, for instance by
$\Delta a'(0)= 2 a'(0^+)$, and a similar relation for $\Delta n'(0)$.
 
Let us notice that if we use the above boundary
conditions, we may  evaluate equation (\ref{G04})  on the brane to get
the conservation equation 
 \be 
 \dot \rho_b +3{\dot a \over a}(p_b+\rho_b)=0.
 \label{drhob}
 \ee
This is a general result that is independent of the bulk content. 
In the subsequent discussion, we will assume that the bulk is stable and  
use the freedom of coordinate redefinition to set $b=1$.
 This condition will simplify  the Einstein equations making it
easier to extract its physical meaning, as we will see later.

\section{Master equations of brane cosmology}
\setcounter{equation}{0}

Once we assume that the fifth dimension is stable, we can follow 
standard procedure as in four dimensional cosmology to reduce the equations
given in the last section to a minimal set. First, notice that the
conservation laws (\ref{drho}), (\ref{dp}) and (\ref{drhob}), 
will have $\dot b = 0$. Taking $b=1$, we notice
that the $G_{00}$ component (Eq. \ref{G00}) of the Einstein equations is
already the equivalent of the Friedmann equation, with the Hubble
parameter defined as a function of $y$. However, the presence of the brane
will require that the term with second derivative be regularized by
extracting the divergent part. The Friedmann equation then has the form
 \be 
  H^2(y):=\left(\frac{\dot a}{a}\right)^2 = 
  \frac{\kappa_5^2}{3} n^2 \rho_B + 
  n^2\left[\left(\frac{a'}{a}\right)^2 + \frac{a''_R}{a}\right] -
  k\frac{n^2}{a^2};
  \label{H}
 \ee  
where $a''_R$ stands for the regular part of the function. Clearly, we
recognize an
expression similar to that given  early~\cite{binetruy2}. However, let us
stress that 
there  is no unknown  constant of integration  as
in~\cite{binetruy2}. Also, we may identify the regular term as the
contribution of the Weyl tensor of the bulk~\cite{shiru}. Since this
expression is continuous, thanks to parity symmetry, we may evaluate it on
the
brane to get the effective Friedmann equation of our universe, where the
time component of the metric tensor is chosen to be $n_0=1$:
 \be 
 H^2_0 = 
  \frac{\kappa_5^2}{3} \rho_{B0} + 
  \left({\kappa_5^2\over 6} \rho_b\right)^2 + 
   \left(\frac{a''_R}{a}\right)_0 -  {k\over a^2_0},
  \label{Heff}
 \ee  
where the subindex $0$ stand for the evaluation at $y=0$.
As already known\cite{binetruy1}, this expression has the
squared dependence
on the brane density. On the other hand, it also has a dependence on
the metric outside (i.e. in the bulk), through $a''_R$, which however
could
be evaluated, once we solve for $a$ as a function of $y$. Particularly,
for simple
cases as an empty bulk, the $y$ dependence can be explicitly extracted
and this term evaluated. 

Next, let us consider the equation involving $G_{ij}$, (\ref{Gij}).
Again, we separate the singular and the regular parts of the
second derivative term and by introducing (\ref{H})  we reduce this
equation to a form equivalent to the acceleration equation given by
 \be
 \frac{\ddot a}{a}=
  -{\kappa_5^2\over 6}\left(3 P_B + \rho_B\right) n^2 + 
  n^2\left(\frac{a'}{a}\frac{n'}{n}\right) + H\frac{\dot n}{n}
  + {n^2\over 2}\left[\frac{a''_R}{a}+\frac{n''_R}{n}\right].
 \label{acc}
 \ee	
Note that, as in the standard FRW cosmology, this is not an independent
equation. Indeed, it can be derived 
by taking the time derivative of (\ref{H}) and combining that expression
with the energy conservation law (\ref{drho}) and $G_{04}$ equation
(\ref{G04}). This derivation holds regardless of whether the bulk radius
is constant or changing with time. As a result,
this equation does not provide any extra
information, but it will be useful in what follows.

We now turn to the equation involving $G_{44}$, (\ref{G44}).  In
conjunction with the equation involving $G_{04}$, this represents
 the new ingredient of the brane cosmology and can be a 
window to understand the $y$ dependence of the cosmological 
parameters. Notice that our procedure is in contrast to that used in
previous works where the Friedmann equation has been obtained from 
$G_{44}$  component. 
If we substitute Eqs. (\ref{H}) and (\ref{acc}) in the equation for the
$G_{44}$ component (Eq. (2.9)), we find a
simple equation that governs the behaviour of $a(y,t)$ and $n(y,t)$ 
on the bulk (into each one of the domains). It is given by
 \be
 3\frac{a''}{a}+\frac{n''}{n}=
 \frac{\kappa_5^2}{3}\left(3 P_B - 2P_T - \rho_B\right).
 \label{a''}
 \ee
This expression is supplemented by $G_{04}$ which trivialy reduces to
\be
n(y,t)=\lambda(t)\dot a(y,t),
\label{na}
\ee
where $\lambda(t)$ is an arbitrary function of time. Using the
freedom of fixing the gauge on the coordinate system, we can set
 $n_0=1$ in which case we get $\lambda= \dot a_0^{-1}$. However in most of
the results this choice is not necessary at all.

Let us emphasize that Eqs. (\ref{H}),  (\ref{a''}) and (\ref{na}) form the
set of master equations, in the sense that they determine the $t$ and $y$
dependence of the metric. We study the solutions of these equations in the
next section.

\section{Exact solutions on the bulk}
\setcounter{equation}{0}

As a simple application of our master equations let us
reconsider the cases  already studied in the literature. 
Let us assume that the stress tensor of the bulk gets contribution only
from a cosmological constant, ${\hat T^A}{_B} =-\Lambda_B \delta^A_B$. 
Then, Eq. (\ref{a''}) reduces into
 \be
  3\frac{a''}{a} + \frac{n''}{n}= - \frac{2}{3}\kappa_5^2\Lambda_B.
  \label{alambda}
 \ee 
This equation, together with the scaling equation (\ref{na}),
can be solved in a straightforward manner. For $\Lambda_B=0$ we get a
linear solution just as in~\cite{binetruy1}
 \be 
 a(y,t)=A |y| + B; \qquad n= \lambda \left(\dot A |y| + \dot B\right).
 \ee
Here we have already imposed $P$ (parity) symmetry on the solution. Next, 
by using the boundary conditions (\ref{gap1}) and (\ref{gap2}) we get 
the final result
 \be
 a(y,t)=a_0\left (1-\frac{\kappa_5^2}{6}\rho_b|y|\right); \qquad 
 \mbox{and} \qquad 
 n(y,t)=n_0\left (1+\frac{\kappa_5^2}{6}(3p_b + 2\rho_b)|y|\right).
 \ee
Notice that in  the expression for $n$ requiring consistency with  the
scaling (\ref{na}) leads to the conservation law 
 \be
  \dot \rho_b + 3H_0 (p_b+\rho_b) = 0.
 \label{drb}
 \ee
Note that $a''_R=0$ in this case, and there is no obvious way to get the
correct Friedmann equation (\ref{Heff}) (i.e. linear rather than squared
dependence of $H^2$ on $\rho_b$).

Next, we  assume that $\Lambda_B$ is non zero. Again
(\ref{alambda}) is easy to solve, and after using the boundary conditions, 
we get for $\Lambda_B<0$
 \bea
 a(y,t) &=& a_0 \left( \cosh(\mu|y|) - 
 \frac{\kappa_5^2}{6\mu} \rho_b \sinh(\mu|y|)\right), \nonumber\\
 n(y,t) &=& n_0 \left( \cosh(\mu|y|) +
  \frac{\kappa_5^2}{6\mu}(3p_b + 2\rho_b)\sinh(\mu|y|)\right); 
  \label{sol}
 \eea
where $\mu^2 = - \kappa_5^2\Lambda_B/6$. We recognize those solution
presented by Binetruy et al. in~\cite{binetruy2} when they take their
integration constant as zero. Our finding is that this is the only allowed
solution
with a stable extra dimension. In this case, the condition (\ref{na})
reduces
to (\ref{drb}). For $\Lambda_B>0$ the solutions have a similar form,
with the hyperbolic functions replaced by $\cos$ and $\sin$ respectively. 

If we also assume a brane tension, $\Lambda_b$, the Friedmann equation on
the brane becomes
 \be 
  H^2_0 =  
  {\kappa_5^4\over 18}\Lambda_b \rho_b + 
  {\kappa_5^4\over 36} \rho_b^2 -  {k\over a^2_0} + 
  {\kappa_5^4\over 36} \Lambda_b^2 + 
  \frac{\kappa_5^2}{6} \Lambda_B .
  \label{H0}
 \ee  
We emphasize that this result shows that ${\cal C}$, the unknown constant
found in~\cite{binetruy2} is actually zero.
A fine tunning among the cosmological constants~\cite{cline}
 \be 
 \Lambda_B  = -\frac{\kappa_5^2}{6} \Lambda_b^2
  \label{tune}
 \ee
may linearize the Friedmann equation in the limit where $\rho_b\ll
\Lambda_b$. As it is clear, this will work only if the vacuum energy of the
bulk is negative. Moreover, to get the right expression of the Friedmann
equation as in standard cosmology, an extra fine tunning~\cite{cline}
 \be 
 \Lambda_b = 6 \frac{\kappa_4^2}{\kappa_5^4}
 \label{tune2}
 \ee
is required, where $\kappa_4^2 = M_{P\ell}^{-2}$. With those new
assumptions the solution (\ref{sol}) reduces into
 \bea
  a = a_0 \left( e^{-\mu|y|} -
   {\rho_b\over \Lambda_b} \sinh(\mu |y|)\right) \nonumber \\
  n = n_0 \left(e^{-\mu|y|} +
   {3p_b + 2\rho_b\over \Lambda_b} \sinh(\mu |y|)\right)
 \eea
Clearly, if we neglect the contribution of the brane densities, we
identify the Randall-Sundrum static solutions~\cite{rs}
 \be
 a=a_0e^{-\mu|y|} \qquad \mbox{and} \qquad n=n_0e^{-\mu|y|}.
 \label{rs}
 \ee

So far we have only been concerned with finding solutions of the
five dimensional cosmological equations with an empty bulk. If we wanted
to include a bulk field, e.g. to provide a new picture of
inflation\cite{binf}, one would have a modified form of the Equation 
\ref{a''}:
 \be 
  3\frac{a''}{a}+\frac{n''}{n}=
 -2\frac{\kappa_5^2}{3}\left(V(\phi) + \Lambda_B\right).
 \ee
Clearly, if the scalar field evolves very slowly, e.g in the inflation
epoch, the solution will
be approximately  of the form (\ref{sol}), but now with 
 \be 
 \mu^2 = - {\kappa_5^2 \over 6} \left( V(\phi) + \Lambda_B\right).
 \ee
By setting the fine tunning (\ref{tune}) on (\ref{Heff}) we get 
 \be 
  H^2_0 =  
   \frac{\kappa_5^2}{6} V(\phi(0)) -  {k\over a^2_0} + 
   {\kappa_5^4\over 18}\Lambda_b \rho_b + 
  {\kappa_5^4\over 36} \rho_b^2 .
 \ee  
As expected, the inflaton potential  contributes linearly to last equation,
and the effective potential is just the value of the bulk potential on the
brane.

\section{Non factorization of the metric parameters}
\setcounter{equation}{0}

Before closing the present discussion,
 let us comment briefly on the nature of the
exact solutions. First we point out that all the solutions presented on the
previous section are not factorizable. This is in fact a general
property of this class of models as we will show now. For this purpose,
let us start with case when the bulk radius $b$ is time dependent and we
will see that factorization of the metric parameters $n(y,t)$ and $a(y,t)$
will not be consistent with a stable $b$. For this purpose, we start with
the identity
 \be 
 {d\over dt}\left({a'\over a}\right) = {\dot a'\over a} - 
 {\dot a \over a}{a'\over a} ;
  \label{da'}
 \ee
which, together with the $G_{04}$ equation  leads to 
 \be 
 {d\over dt}\left({a'\over a}\right) = 
 H \left({n'\over n} - {a'\over a} \right) + 
 {\dot b \over b}{a'\over a} .
  \label{daa}
 \ee
Note that this expression reduces to Eq. (\ref{drhob}) on the brane. 

Now, let us assume that $a$ and $n$ are factorizable i.e. 
 \be
 a = a_0(t) \beta_1(y) \qquad n=n_0(t)\beta_2(y). 
 \ee 
One can then easily see that
 \be 
 {d\over dt}\left({a'\over a}\right) = 0.
 \label{daa2}
 \ee
This can be used to rewrite Eq. (\ref{daa}) as
 \be
  \frac{\dot b}{b} + H_0\left( 
  \frac{ \beta_1 \beta'_2} {\beta'_1 \beta_2} - 1 \right) = 0.
 \label{betas}
 \ee
This equation can be integrated in a straightforward manner and leads to  
 \be
 b = a_0^{-\beta_3(y)};
 \ee
where $\beta_3$ is given in terms of $\beta_{1,2}$ by the expression
between parenthesis in (\ref{betas}).
From this, we conclude that the factorization ansatz works when the
bulk radius, $b$ is time dependent since clearly $a_0$ grows with time.
A stable $b$ would then mean that
$\beta_3$ must equal zero, which is possible only if $\beta_1=\beta_2$.
Furthermore, since Eq. (\ref{daa2}) holds everywhere,   we may
evaluate it on the brane and then conclude using Eq. (2.12) that 
$\dot\rho_b = 0$. From this, it follows that factorizability of $n(y,t)$ and $a(y,t)$
proposed above implies that only a cosmological constant may be present
in the brane. Since we require a true time dependent density in the
brane to describe realistic cosmology e.g. the transition from a radiation
to matter dominated universe, the exact solution to the metric 
can not be factorizable.

Parenthetically, let us note that if we had chosen a
 more general form for the exact solutions
 \be 
 a = a_0(t) \beta_1(y,t) \qquad n=n_0(t)\beta_2(y,t);
 \ee 
where the functions $\beta_{1,2}$ satisfy the boundary condition
$\beta_1(0,t)= \beta_2(0,t) = 1$, no such restriction on brane energy
density would emerge. This is indeed the form of the solutions
discussed above. For this case, the five dimensional  scalar curvature
can be written as
 \be 
 R = \beta_2^{-2} R_{(4)} + \dots;
 \ee
where $R_{(4)}$ is the four dimensional scalar curvature formed by $a_0$
and $n_0$, the dots represent  
extra terms.

At this point, we note a puzzling feature with regard to the true 
definition of the Newton's constant. One can define the Newton's constant
in two ways: one by identifying the coefficient in front of the energy
density in Friedmann equation and another way is by integrating over the
fifth dimension in the action integral. We call the first definition a
``local'' one whereas the second one we call ``global''. Clearly in the
presence of matter, the first one gives a constant $G_N$
whereas the second gives a time-dependent $G_N$. We can check this easily
as follows. Consider the gravity action in five dimensions
 \be
 S = \int \! d^4x dy\  \sqrt{-g}{1\over 2\kappa_5^2} R ,
 \ee
Using the relation 
 \be 
 \sqrt{-g} = \sqrt{-g_4}~ \beta_1^3 \beta_2
 \ee
(with $g_4$ the determinant of the corresponding four dimensional metric)
and Eq. (5.8), we obtain
 \be
 {1\over \kappa_4^2} = {1\over \kappa_5^2} \int\! dy\ 
 \beta_1^3 \beta_2^{-1}
 \ee
adopting the ``global'' definition of Newton's constant.
The integral in the last equation is taken over the whole fifth dimension.
It is
clear from the above expression, that this leads, in general to a time
dependent Newton's constant whereas if we used Friedmann equation, 
we would get it to be time independent. We wish to note that
when one uses the static solution,
a l\'a Randall and Sundrum, there is no time dependence due to the absence
of matter and hence no puzzle.


\section{conclusions}
 
In summary, we have analyzed Einstein equations for cosmology  
in five dimensions  within a brane-bulk picture. Restricting to the case
of a
stable bulk radius, we extract the generalized Friedmann
equation which does not contain any integration constant, but a term that
involves a regular part of the second derivative over the spatial
component of the  metric along the brane coordinates. We then evaluate
this second derivative using the complete set of the master equations for
five dimensional
cosmology and find that for the case of a stable bulk radius, there is no
arbitrary constant in the Friedmann equation in the brane. This makes it
easier to interpret the brane cosmology as the standard big bang picture.
An advantage of our analysis is that we do not need to make
any explicit assumptions regarding the bulk content.
In this sense it generalizes the results presented
before in the literature where only a 
cosmological constant was assumed to be present in the bulk to
obtain the Friedmann equation.
Finally we make some remarks on the general nonfactorizable nature of the
exact solutions.


{\it Acknowledgements.}   The work of RNM is supported by a grant from the
National Science Foundation under grant number PHY-9802551. The work of
APL is supported in part by CONACyT (M\'exico). The work of CP is
supported by Funda\c c\~ao de Amparo \`a Pesquisa do Estado de S\~ao Paulo
(FAPESP). We wish to thank C. Van de Bruck and S. Pastor for discussions.


\end{document}